\documentclass[11pt]{article}
\usepackage{amssymb}
\usepackage{amsmath}
\newcommand\doingARLO[2][]{%
  \ifx\mmref\undefined #1\else #2\fi
}
\newcommand{\A}[1]{A^{(#1)}}
\newcommand{\C}[1]{C^{(#1)}}
\newcommand{\G}[1]{G^{(#1)}}
\newcommand{\CHI}[1]{\chi^{(#1)}}

\newcommand{\LAM}[1]{\Lambda^{(#1)}}
\newcommand{\PSI}[1]{\Psi^{(#1)}}

\newcommand{\Int}{\mathop{\rm Int}\nolimits}
\newcommand{\nuv}{\underline{\phantom{\mu }\!\!\!\nu}}
\newcommand{\wdg}{{\scriptscriptstyle \wedge}}
\newcommand{\tNS}{}
\def\rmi{{\rm i}}

\begin{document}
\begin{titlepage}
\begin{flushright}
  UG-01-28 \\
  SU-ITP-01/23 \\
  IFT-UAM/CSIC-01-14 \\
  KUL-TF-01/12 \\
  hep-th/0105061
\end{flushright}
\begin{center}
\vspace{.5cm}\baselineskip=16pt {\LARGE \bf Brane plus Bulk Supersymmetry in Ten Dimensions} \\
\vskip 1 cm {\Large
  Eric Bergshoeff$^1$, Renata Kallosh$^2$, Tom\'as Ort\'{\i}n$^3$, \\[3mm]
Diederik Roest$^1$, and Antoine Van Proeyen$^4$
} \\
\vskip 1 cm
{\small
  $^1$ Institute for Theoretical Physics, Nijenborgh 4, 9747 AG Groningen, The Netherlands \\[3mm]
  $^2$ Department of Physics, Stanford University, Stanford, California 94305,
USA
\\[3mm]
  $^3$ Instituto de F\'{\i}sica Te\'orica, C-XVI, Universidad Aut\'onoma de Madrid, E-28049-Madrid, Spain \\
       I.M.A.F.F., C.S.I.C., Calle de Serrano 113 bis, E-28006-Madrid, Spain \\[3mm]
  $^4$ Instituut voor Theoretische Fysica, Katholieke Universiteit Leuven,\\
       Celestijnenlaan 200D B-3001 Leuven, Belgium
}
\end{center}
\centerline{ABSTRACT}
\bigskip
We discuss a generalized form of  IIA/IIB supergravity depending on {\it
all} R-R potentials $\C{p}\ (p = 0, 1, \ldots 9)$ as the effective field
theory of Type~IIA/IIB superstring theory. For the IIA case we explicitly
break this R-R democracy to either $p\leq3$ or $p\geq5$ which allows us
to write a new bulk action that can be coupled to $N=1$ supersymmetric
brane actions.
\par
The case of 8-branes is studied in detail using the new bulk \& brane
action. The supersymmetric negative tension branes without matter
excitations can be viewed as orientifolds in the effective action. These
D$8$-branes and O$8$-planes are fundamental in Type ${\rm I}^\prime$
string theory. A BPS 8-brane solution is given which satisfies the jump
conditions on the wall. As an application of our results we derive
a quantization of the mass parameter and the cosmological constant
in string units. 
\end{titlepage}

\section*{\begin{center} \Large MOTIVATION \end{center}}


Our purpose is to construct supersymmetric domain
walls of string theory in $D=10$ which may shed some light on the stringy
origin of the brane world scenarios. In the process of pursuing this goal
we have realized that all descriptions of the  effective field theory of
Type~IIA/B string theory available in the literature are inefficient for
our purpose. This has led us to introduce new versions of the effective
supergravities corresponding to Type~IIA/B string theory.

The standard IIA massless supergravity includes the $\C{1}$ and $\C{3}$
R-R potentials and the corresponding $\G{2}$ and  $\G{4}$ gauge-invariant
R-R forms. Type IIB supergravity includes the $\C{0}$, $ \C{2}$ and
$\C{4}$  R-R potentials and the corresponding $\G{1}$, $\G{3}$ and
(self-dual) $\G{5}$  gauge-invariant R-R forms. On the other hand, string
theory has all D$p$-branes, odd and even, including the exotic ones, like
8-branes in IIA and 7-branes in IIB theory. These branes, of co-dimension
1 and 2,  are special objects which are different in many respects from
the other BPS-extended objects like the $p$-branes with $0\le p \le 6$,
which have co-dimension greater than or equal to~3. The basic difference
is in the behavior of the form fields at large distance, $\G{p+2}\sim
r^{p-8}$. For example the $\G{10}$ R-R form of the 8-brane does not fall
off at infinity but takes a constant value there. It is believed that such
extended objects can not exist independently but only in connection with
orientifold planes~\cite{berg-PolchinskiBook}. However, the realization of the
total system in supergravity is rather obscure.

It has been realized a while ago~\cite{berg-Polchinski:1995mt}  that  massive
IIA supergravity, discovered by Romans~\cite{berg-Ro86}, was the key to
understand the spacetime picture of the 8-branes, which are domain walls
in $D=10$. A significant progress towards the understanding of the
8-brane solutions was made 
in~\cite{berg-Polchinski:1996df,berg-Bergshoeff:1996ui},
where the bulk supergravity solution was found. Also,
in~\cite{berg-Bergshoeff:1996ui}, the description of the cosmological constant
via a 9-form potential, based upon the work
of~\cite{berg-Duff:1980qv,berg-Aurilia:1980xj}, was discussed.
In~\cite{berg-Alonso-Alberca:2000ne} a standard 8-brane action coupling to this
9-form potential has been shown to be the appropriate source for the
second Randall--Sundrum scenario~\cite{berg-RS}. Solutions for the coupled bulk
\& brane action system automatically satisfy the jump conditions and so
they are consistent, at least from this point of view. A major unsolved
problem was to find an explicitly supersymmetric description of coupled
bulk \& brane systems like it was done in~\cite{berg-Bergshoeff:2000zn}. Such
a description should allow us to find out some important properties of
the domain walls like the distance between the planes, the status of
unbroken supersymmetry in the bulk and on the brane  etc. We expect that
realizing such a bulk \& brane  construction will lead to a better
insight into the  fundamental nature of extended objects of string theory.
\par
The string backgrounds that we want to describe using an explicitly
supersymmetric bulk \& brane action are one-dimensional orbifolds
obtained by modding out the circle $S^{1}$ by a reflection
$\mathbb{Z}_{2}$. The orbifold direction is the transverse direction of
the branes that fill the rest of the spacetime. Now, the orbifold
$S^{1}/\mathbb{Z}_{2}$ being a compact space, we cannot place a single
charged object (a D$8$-brane, say) in it, but we have to have at least
two oppositely charged objects. However, this kind of system cannot be in
supersymmetric equilibrium unless their tensions also have opposite
signs. We are going to identify these negative-tension objects with
O$8$-planes and we will propose an O$8$-plane action to be coupled to the
bulk supergravity action. O$8$-planes can only sit at orbifold points
because they require the spacetime to be mirror symmetric in their
transverse direction and, thus, they can sit in any of the two endpoints
of the segment $S^{1}/\mathbb{Z}_{2}$. We are going to place the other
(positive tension, opposite R-R charge) brane at the other endpoint.
Clearly we can, from the effective action point of view, identify the
positive tension brane as a combination of O$8$-planes and D$8$-branes
with positive total tension and the negative tension brane as a
combination of O$8$-planes and D$8$-branes with negative total tension.
\par
Our strategy will be to generalize the 5-dimensional construction of the
supersymmetric bulk \& brane action, proposed
in~\cite{berg-Bergshoeff:2000zn}. The construction of~\cite{berg-Bergshoeff:2000zn}
allowed to find a supersymmetric realization of the brane-world scenario
of Randall and Sundrum~\cite{berg-RS}. We will repeat the construction
of~\cite{berg-Bergshoeff:2000zn} in $D=10$ with the aim to get a better
understanding of branes and planes in string theory.

To solve the discrepancy between the bulk actions with limited field
content (lower-rank R-R forms) and the wide range of brane actions that
involve all the possible R-R forms, we have constructed a new formulation
of IIA/IIB supergravity up to quartic order in fermions. In particular,
the new formulation gives an easy control over the exotic $\G{0}$ and
$\G{10}$ R-R forms associated with the mass and cosmological
constant of the $D=10$ supergravity. This in turn allows a clear study of the
D$8$--O$8$ system describing a pair of supersymmetric domain walls which
are fundamental objects of the Type~${\rm I^\prime}$ string theory. The
quantization of the mass parameter and cosmological constant in stringy
units are simple consequences of the theory. Apart from being a tool to
understand the supersymmetric domain walls we were interested in, it can
be expected that the new effective theories of $D=10$ supersymmetry will
have more general applications in the future.

In this talk we will summarize the results of \cite{berg-Bergshoeff:2001pv}.

\section*{\begin{center} \Large A NEW DUAL FORMULATION OF D=10 SUPERGRAVITY \end{center}}

The standard formulation of $D=10$ IIA
(massless~\cite{berg-Huq:1985im,berg-Giani:1984wc,berg-Campbell:1984zc} and
massive~\cite{berg-Ro86}) and IIB~\cite{berg-Schwarz:1983qr,berg-Howe:1984sr}
supergravity has the following field content
\begin{eqnarray}
{\rm IIA}&:&\hskip 1truecm  \left\{
  g_{\mu \nu},
  B_{\mu \nu},
  \phi,
  \C{1}_{\mu},
  \C{3}_{\mu \nu \rho},
  \psi_\mu,
  \lambda
  \right\} \,, \cr
{\rm IIB}&:&\hskip 1truecm  \left\{
  g_{\mu \nu},
  B_{\mu \nu},
  \phi,
  \C{0},
  \C{2}_{\mu \nu},
  \C{4}_{\mu\nu\rho\sigma},
  \psi_\mu,
  \lambda
  \right\} \, .
\label{berg-convfc}
\end{eqnarray}
In the IIA case, the massive theory contains an additional mass parameter
$G^{(0)} = m$. In the IIB case, an extra self-duality condition is imposed
on the field strength of the four-form. It turns out that one can realize
the N=2 supersymmetry on the R-R gauge fields of higher rank as well.
These are usually incorporated via duality relations. To treat the R-R
potentials democratically we propose a new
formulation based upon a pseudo-action. This democratic formulation
describes the dynamics of the bulk supergravity in the most elegant way.
However, it turns out that this formulation is not well suited for our
purposes. For the IIA case, we therefore give a different formulation 
where the constant mass parameter has been
replaced by a field.

\section*{\begin{center} The Democratic Formulation \end{center}}
\label{berg-ss:Dformulation} 

To explicitly introduce the democracy among the
R-R potentials we propose a pseudo-action whose equations of motion 
are supplemented by duality constraints (see below). Of
course this enlarges the number of degrees of freedom. Since a $p$- and
an $(8-p)$-form potential carry the same number of degrees of freedom,
the introduction of the dual potentials doubles the R-R sector. Including
the highest potential $\C{9}$ in IIA does not alter this, since it carries
no degrees of freedom. This 9-form potential can be seen as the potential
dual to the constant mass parameter $G^{(0)} = m$. The doubling of number
of degrees of freedom will be taken care of by a constraint, relating the
lower- and higher-rank potentials. This new formulation of supersymmetry
is inspired by the bosonic construction of~\cite{berg-Fukuma:1999jt}, and, in
the case of IIB supergravity, is related to the pseudo-action construction
of~\cite{berg-Bergshoeff:1996sq}.
\par
A pseudo-action~\cite{berg-Bergshoeff:1996sq} can be used as a mnemonic to
derive the equations of motion. It differs from a usual action in the
sense that not all equations of motion follow from varying the fields in
the pseudo-action. To obtain the complete set of equations of motion, an
additional constraint has to be substituted by hand into the set of
equations of motion that follow from the pseudo-action. The constraint
itself does not follow from the pseudo-action. The construction we
present here generalizes the pseudo-action construction
of~\cite{berg-Fukuma:1999jt,berg-Bergshoeff:1996sq} in the sense that our
construction (i) treats the IIA and IIB case in a unified way,
introducing all R-R potentials in the pseudo-action, and (ii) describes
also the massive IIA case via a 9-form potential $C^{(9)}$ and a constant
mass parameter $G^{(0)} = m$.
\par
Our pseudo-action has the extended field content
\begin{eqnarray}
{\rm IIA}&:&\hskip 1truecm  \left\{
  g_{\mu \nu},
  B_{\mu \nu},
  \phi,
  \C{1}_{\mu},
  \C{3}_{\mu \nu \rho},
  \C{5}_{\mu \cdots \rho},
  \C{7}_{\mu \cdots \rho},
  \C{9}_{\mu \cdots \rho},
  \psi_\mu,
  \lambda
  \right\} \, , \cr
{\rm IIB}&:&\hskip 1truecm  \left\{
  g_{\mu \nu},
  B_{\mu \nu},
  \phi,
  \C{0},
  \C{2}_{\mu \nu},
  \C{4}_{\mu \cdots \rho},
  \C{6}_{\mu \cdots \rho},
  \C{8}_{\mu \cdots \rho},
  \psi_\mu,
  \lambda
  \right\} \, .
\label{berg-fcdemo}
\end{eqnarray}
It is understood that in the IIA case the fermions contain both
chiralities, while in the IIB case they satisfy
\begin{align}
  \Gamma _{11} \psi_\mu = \psi_\mu\,, \qquad \Gamma _{11} \lambda = -
  \lambda\,, \qquad
  \text{(IIB).}
\label{berg-chirality}
\end{align}
In that case they are doublets, and we suppress the corresponding index.
The explicit form of the pseudo-action is given by\footnote {We use
the notation and conventions of \cite{berg-Bergshoeff:2001pv}.}
\begin{align}
 S_{\text{Pseudo}} =
  & - \frac{1}{2\kappa_{10}^2}\int d^{10} x \sqrt{-g}
    \Big\{
    e^{-2\phi} \Big[
    R\big(\omega(e)\big) -4\big( \partial{\phi} \big)^{2}
    +\tfrac{1}{2} H \cdot H + \nonumber\\
 &  \hspace{1cm}  -2\partial^\mu\phi \CHI{1}_{{\mu}}
        + H \cdot \CHI{3}
   +2 \bar{{\psi}}_{{\mu}}{\Gamma}^{{\mu}{\nu}{\rho}}
    {\nabla}_{{\nu}}{\psi}_{{\rho}}
    -2 \bar{{\lambda}}{\Gamma}^{{\mu}}
    {\nabla}_{{\mu}}{\lambda}
    +4 \bar{{\lambda}} {\Gamma}^{{\mu}{\nu}}
    {\nabla}_{{\mu}}{\psi}_{{\nu}}
    \Big]+ \displaybreak[2] \notag \\
  & \hspace{1cm} + \sum_{n=0,1/2}^{5,9/2}
    \tfrac{1}{4} \G{2n} \cdot \G{2n}
    + \tfrac{1}{2} \G{2n} \cdot \PSI{2n} \Big\}+ \mbox{ quartic fermionic terms}\, .
    \label{berg-IIABactiondemo}
\end{align}
It is understood that the summation in the above pseudo-action is over
integers ($n=0,1,\ldots ,5$) in the IIA case and over half-integers
($n=1/2,3/2,\ldots ,9/2$) in the IIB case. In the summation range we will
always first indicate the lowest value for the IIA case, before the one
for the IIB case. Furthermore,
\begin{equation}
\frac{1}{2\kappa_{10}^2}= \frac{g^2}{2\kappa^2}=\frac{2\pi}{(2\pi
\ell_s)^8}\,, \label{berg-coupling}
\end{equation}
where $\kappa^2$ is the physical gravitational coupling, $g$ is  the
string coupling constant and $\ell_s= \sqrt {\alpha '}$ is the string
length. For notational convenience we group all potentials and field
strengths in the formal sums
\begin{align}
  {\bf G} = \, \sum_{n=0,1/2}^{5,9/2} \G{2n} \,, \hspace{1.5cm}
  {\bf C} = \, \sum_{n=1,1/2}^{5,9/2} \C{2n-1} \,. \label{berg-formsums}
\end{align}
The bosonic field strengths are given by
\begin{align}
  H = d B \,, \qquad
  {\bf G} = d {\bf C} - d B \wdg {\bf C} + \G{0} {\bf e}^B \,,
\label{berg-G2n}
\end{align}
where it is understood that each equation involves only one term from the
formal sums (\ref{berg-formsums}) (only the relevant combinations are
extracted). The corresponding Bianchi identities then read
\begin{align}
  d H =0 \,, \hspace{1cm}
  d {\bf G} - H \wdg {\bf G} =0 \,.
\label{berg-Bianchis}
\end{align}
In this subsection $G^{(0)} =m$ indicates the constant mass parameter of
IIA supergravity. In the IIB theory all equations should be read with
vanishing $\G{0}$. The spin connection in the covariant derivative
$\nabla_\mu $ is given by its zehnbein part: $\omega_{\mu}^{\; \;
ab}=\omega_{\mu}^{\; \; ab}(e)$. The bosonic fields couple to the
fermions via the bilinears $\CHI{1,3}$ and $\PSI{2n}$, which read
\begin{align}
  \CHI{1}_\mu =
  & -2 \bar{\psi}_\nu \Gamma^\nu \psi_\mu
    -2 \bar{\lambda} \Gamma^\nu \Gamma_\mu \psi_\nu\, , \notag \\
  \CHI{3}_{\mu\nu\rho} =
  & \tfrac{1}{2}\bar{{\psi}}_{{\alpha}}
    {\Gamma}^{[{\alpha}}
    \Gamma_{\mu\nu\rho}
    {\Gamma}^{{\beta}]}
    {\cal P}{\psi}_{{\beta}}
    + \bar{{\lambda}}
    \Gamma_{\mu\nu\rho}{}^{\beta}
    {{\cal P}}{\psi}_{{\beta}}
    -\tfrac{1}{2}\bar{{\lambda}} {\cal P}\Gamma_{\mu\nu\rho}
    {\lambda}\, , \notag \\
  \PSI{2n}_{\mu_1\cdots \mu_{2n}} =
  & {\textstyle\frac{1}{2}}e^{-{\phi}}
    \bar{{\psi}}_{{\alpha}}
    {\Gamma}^{[{\alpha}}
    \Gamma_{\mu_1\cdots \mu_{2n}}
    {\Gamma}^{{\beta}]}
    {\cal P}_n {\psi}_{{\beta}}
    +{\textstyle\frac{1}{2}}e^{-{\phi}}
    \bar{{\lambda}}
    \Gamma_{\mu_1\cdots \mu_{2n}}
    {\Gamma}^{{\beta}}
    {\cal P}_n{\psi}_{{\beta}}+ \notag \\
  & -{\textstyle\frac{1}{4}}
    e^{-{\phi}}
    \bar{{\lambda}}
    \Gamma_{ [ \mu_1\cdots \mu_{2n-1}}
    {\cal P}_n \Gamma_{\mu_{2n} ] } {\lambda}\, .
\label{berg-fermionbilinears}
\end{align}
We have used the following definitions:
\begin{eqnarray}
{\cal P} &=& \Gamma_{11}\ \ \quad {\rm (IIA)} \qquad {\rm or}\qquad   -\sigma^3\ {\rm (IIB)}\,, \nonumber\\
{\cal P}_n &=& (\Gamma_{11})^{n}\ {\rm (IIA)}\qquad  {\rm or}\qquad
\sigma^1\ ({\rm n+1/2\ even}),\ \rmi\sigma^2\ ({\rm n+1/2\ odd}) \ ({\rm
IIB})\,.
\end{eqnarray}
Note that the fermions satisfy
\begin{equation}
\PSI{2n} = (-)^{\Int[n]+1} \star \PSI{10-2n}\,.
\label{berg-Psidual}
\end{equation}
\par
Due to the appearance of all R-R potentials, the number of degrees of
freedom in the R-R sector has been doubled. Each R-R potential leads to a
corresponding equation of motion:
\begin{align}
 d \star (\G{2n}+\PSI{2n})
  + H \wdg \star (\G{2n+2}+\PSI{2n+2}) = 0\, . \label{berg-eoms}
\end{align}
Now, one must relate the different potentials to get the correct number of
degrees of freedom. We therefore by hand impose the following duality
relations
\begin{align}
  \G{2n} + \PSI{2n} = (-)^{\Int[n]} \star \G{10-2n}\, ,
\label{berg-duality}
\end{align}
in the equations of motion that follow from the pseudo-action
(\ref{berg-IIABactiondemo}). It is in this sense that the action
(\ref{berg-IIABactiondemo}) cannot be considered as a true action. Instead, it
should be considered as a mnemonic to obtain the full equations of motion
of the theory. As usual, the Bianchi identities and equations of motions
of the dual potentials correspond to each other when employing the
duality relation. For the above reason the democratic formulation can be
viewed as self-dual, since (\ref{berg-duality}) places constraints relating
the field content (\ref{berg-fcdemo}).
\par
The pseudo-action (\ref{berg-IIABactiondemo}) is invariant under supersymmetry
provided we impose the duality relations (\ref{berg-duality}) after varying
the action. The supersymmetry rules read (here given modulo cubic fermion
terms):
\begin{align}
  \delta_{{\epsilon}} {e}_{{\mu}}{}^{{a}} =
  & \bar{{\epsilon}}{\Gamma}^{{a}} {\psi}_{{\mu}}\, , \notag \\
  \delta_{{\epsilon}} {\psi}_{{\mu}} =
  & \Big( \partial_{{\mu}} +\tfrac{1}{4}
    \not\!{\omega}_{{\mu}}
    +\tfrac{1}{8}{\cal P}\not\!\! {H}_{\mu}
    \Big) {\epsilon}
    +\tfrac{1}{16} e^{{\phi}} \sum_{n=0,1/2}^{5,9/2} \frac{1}{(2n)!}
    \not \! {G}^{(2n)} {\Gamma}_{{\mu}}
    {\cal P}_n{\epsilon}\, , \notag \\
  \delta_{{\epsilon}} B_{\mu\nu} =
  & -2 \, \bar{{\epsilon}} \Gamma_{[\mu}
    {\cal P} {\psi}_{\nu]}\, , \notag \\
  \delta_\epsilon \C{2n-1}_{\mu_1\cdots \mu_{2n-1}} =
   & - e^{-\phi} \, \bar{\epsilon} \,
    \Gamma_{[\mu_1\cdots \mu_{2n-2}} \, {\cal P}_n \,
\Big((2n-1) \psi_{\mu_{2n-1}]} - \tfrac{1}{2} \Gamma_{\mu_{2n-1}]}
\lambda\Big)
    + \notag \\
    & +(n-1)(2n-1) \, \C{2n-3}_{[\mu_1\cdots \mu_{2n-3}} \,
   \delta_\epsilon B_{\mu_{2n-2}\mu_{2n-1}]}\, , \displaybreak[2] \notag \\
  \delta_{{\epsilon}}{\lambda} =
  & \Big( \! \! \not \! \partial \phi
    + \tfrac{1}{12} \not\!\! {H} {\cal P}\Big) {\epsilon}
    + \tfrac{1}{8} e^{{\phi}} \sum_{n=0,1/2}^{5,9/2} (-)^{2n} \frac{5-2n}{(2n)!}
   \not\! {G}^{(2n)} {\cal P}_n
    {\epsilon}\, , \notag \\
  \delta_{{\epsilon}}{\phi} =
  & \tfrac{1}{2} \, \bar{{\epsilon}}{\lambda}\,,
\label{berg-IIABsusydemo}
\end{align}
where $\epsilon $ is a spinor similar to $\psi _\mu $, i.e. in IIB:
$\Gamma _{11}\epsilon =\epsilon$. Note that for $n$ half-integer (the IIB
case) these supersymmetry rules exactly reproduce the rules given in
eq.~(1.1) of~\cite{berg-BRJO99}.
\par
Secondly, the pseudo-action (\ref{berg-IIABactiondemo}) is also invariant under
the usual bosonic NS-NS and R-R gauge symmetries with parameters
$\Lambda^{\tNS}$ and $\LAM{2n}$ respectively:
\begin{align}
  \delta_\Lambda B =  \, d \Lambda^{\tNS} \,, \qquad
  \delta_\Lambda {\bf C} = \, (d {\bf L}
     - \G{0} \Lambda^{\tNS}) \wdg {\bf e}^B \,, \qquad
  \text{with~~} {\bf L} = \sum_{n=0,1/2}^{4,7/2} \LAM{2n} \,.
\label{berg-gaugedemo}
\end{align}
\par
Finally, there is a number of $\mathbb{Z}_2$-symmetries. However, in the
IIA case these $\mathbb{Z}_2$-symmetries are {\it only valid for $G^{(0)}
= m =0$}. Below we show how these symmetries of the
action act on supergravity fields. For both massless IIA and IIB there is
a fermion number symmetry $(-)^{F_L}$ given by
\begin{align}
  \big\{ \phi, g_{\mu \nu}, B_{\mu \nu} \big\}
  & \rightarrow \big\{ \phi, g_{\mu \nu}, B_{\mu \nu} \big\}\, , \notag \\
  \big\{\C{2n-1}_{\mu_1 \cdots \mu_{2n-1}} \big\}
  & \rightarrow - \big\{ \C{2n-1}_{\mu_1 \cdots \mu_{2n-1}} \big\} \, , \notag \\
  \big\{ \psi_\mu , \lambda, \epsilon \big\}
  & \rightarrow  + {\cal P}
    \big\{ \psi_\mu , - \lambda, \epsilon \big\}\, , \text{~~(IIA),} \notag \\
  \big\{ \psi_\mu , \lambda, \epsilon \big\}
  & \rightarrow  + {\cal P}
    \big\{ \psi_\mu , \lambda, \epsilon \big\}\, , \hskip .4truecm
\text{~~(IIB).}
\label{berg-Omegahat}
\end{align}
In the IIB case there is an additional worldsheet parity symmetry
$\Omega$ given by
\begin{align}
  \big\{ \phi, g_{\mu \nu}, B_{\mu \nu} \big\}
  & \rightarrow \big\{ \phi, g_{\mu \nu}, - B_{\mu \nu} \big\}\, , \notag \\
  \big\{\C{2n-1}_{\mu_1 \cdots \mu_{2n-1}} \big\}
  & \rightarrow (-)^{n+1/2}\big\{ \C{2n-1}_{\mu_1 \cdots \mu_{2n-1}} \big\} \, , \notag \\
  \big\{ \psi_\mu , \lambda, \epsilon \big\}
  & \rightarrow \sigma^1
    \big\{ \psi_\mu , \lambda, \epsilon \big\}\, ,
\label{berg-Omega}
\end{align}
In the massless IIA case there is a similar $I_9\Omega$-symmetry involving
an additional parity transformation in the 9-direction. Writing $\mu =
(\underline \mu, \dot 9)$, the rules are given by
\begin{align}
  x^{\dot 9} & \rightarrow -x^{\dot 9}\, , \notag \\
  \big\{ \phi, g_{\underline{\mu} \nuv },
     B_{\underline{\mu} \nuv} \big\}
  & \rightarrow \big\{ \phi, g_{\underline{\mu} \nuv},
    -B_{\underline{\mu} \nuv} \big\}\, , \notag \\
  \big\{ \C{2n-1}_{\underline{\mu_1} \cdots \underline{\mu_{2n-1}}} \big\}
  & \rightarrow (-)^{n+1} \big\{
    \C{2n-1}_{\underline{\mu_1} \cdots \underline{\mu_{2n-1}}} \big\}
    \,, \notag \\
  \big\{ \psi_{\underline{\mu}} , \lambda, \epsilon \big\}
  & \rightarrow +  \Gamma^{9} \big\{ \psi_{\underline{\mu}} ,
   -\lambda, \epsilon \big\}\,.
\label{berg-I9Omega}
\end{align}
The parity of the fields with one or more indices in the
$\dot{9}$-direction is given by the rule that every index in the
$\dot{9}$-direction gives an extra minus sign compared to the above rules.
\par
In both IIA and IIB there is also the obvious symmetry of interchanging
all fermions by minus the fermions, leaving the bosons invariant.
\par
The $\mathbb{Z}_2$-symmetries are used for the construction of superstring
theories with sixteen supercharges, see~\cite{berg-Bergshoeff:2000re}.
$(-)^{F_L}$ gives a projection to the $E_8 \times E_8$ heterotic
superstring (IIA) or the SO(32) heterotic superstring theory (IIB).
$\Omega$ is used to reduce the IIB theory to the SO(32) Type I
superstring, while the $I_9\Omega$-symmetry reduces the IIA theory to the
${\rm Type\ I}^\prime\, SO(16) \times SO(16)$ superstring theory.
\par
One might wonder at the advantages of the generalized pseudo-action
(\ref{berg-IIABactiondemo}) above the standard supergravity formulation. At the
cost of an extra duality relation we were able to realize the R-R
democracy in the action. Note that only kinetic terms are present; by
allowing for a larger field content the Chern--Simons term is eliminated.
Under T-duality all kinetic terms are easily seen to transform into each
other~\cite{berg-Meessen:1998qm}. The same goes for the duality constraints.
This formulation is elegant and comprises all potentials. However, it is
impossible to construct a proper action in this formulation due to the
doubling of the degrees of freedom. Therefore, to add brane actions to
the bulk system, the democratic formulation is not suitable. This is due
to two reasons. First, the $I_9\Omega$ symmetry is only valid for
$G^{(0)} = 0$, but we will need this symmetry in our construction of the
bulk \& 8-brane system. Secondly, to describe a charged domain wall, we
would like to have opposite values for $G^{(0)}$ at the two sides of the
domain wall, i.e.~we want to allow for a mass parameter that is only
piecewise constant. The R-R democracy has to be broken to accommodate for
an action and this will be discussed in the next subsection.

\section*{\begin{center} The Dual Formulation \end{center}}
\label{berg-ss:action}

We will present here the new dual formulation with action, available for
the IIA case only. A proper action will be constructed in this
formulation. It is this formulation that we will apply in our
construction of the bulk \& brane system. We will call this the dual
formulation.
\par
The independent fields in this formulation are
\begin{equation}
  \left\{   e_\mu ^a,
  B_{\mu \nu},
  \phi,
  \G{0},
  \G{2}_{\mu \nu },
  \G{4}_{\mu _1\cdots \mu _4},
  \A{5}_{\mu _1\cdots\mu _5 },
  \A{7}_{\mu _1\cdots\mu _7 },
  \A{9}_{\mu _1\cdots\mu _9 },
  \psi_\mu,
  \lambda
  \right\}.
 \label{berg-indepFields}
\end{equation}
The bulk action reads
\begin{align}
 S_{\text{bulk}} =
  & - \frac{1}{2\kappa_{10}^2}\int d^{10} x \sqrt{-g}
    \Big\{
    e^{-2\phi} \big[
    R\big(\omega(e)\big) -4\big( \partial{\phi} \big)^{2}
    +\tfrac{1}{2} H \cdot H
    -2\partial^{{\mu}}{\phi} \CHI{1}_{{\mu}}
        + H \cdot \CHI{3}+ \notag \\
  & +2 \bar{{\psi}}_{{\mu}}{\Gamma}^{{\mu}{\nu}{\rho}}
    {\nabla}_{{\nu}}{\psi}_{{\rho}}
    -2 \bar{{\lambda}}{\Gamma}^{{\mu}}
    {\nabla}_{{\mu}}{\lambda}
    +4 \bar{{\lambda}} {\Gamma}^{{\mu}{\nu}}
    {\nabla}_{{\mu}}{\psi}_{{\nu}}
    \big]
    +   \sum_{n=0,1,2}
     \tfrac{1}{2} \G{2n} \cdot \G{2n}
    + \G{2n} \cdot \PSI{2n}  +  \notag \\
  & - \star \, \big[
      \tfrac{1}{2} \, \G{4} \G{4} B
    - \tfrac{1}{2} \, \G{2} \G{4} B^2
    + \tfrac{1}{6} \, \G{2}{}^2 B^3
    + \tfrac{1}{6} \, \G{0} \G{4} B^3 -\tfrac{1}{8} \, \G{0} \G{2} B^4+ \notag \\
  & +\tfrac{1}{40} \, \G{0}{}^2 B^5
    + {\bf e}^{- B} {\bf G}
    d (\A{5} - \A{7} + \A{9}) \big] \Big\}  +\mbox{ quartic fermionic terms}\,,
    \label{berg-dualaction}
\end{align}
where all $\wdg$'s have been omitted in the last two lines. In the last
term a projection on the 10-form is understood.  Here $\textbf{G}$ is
defined as in (\ref{berg-formsums}) but where $\G{0}$, $\G{2}$ and $\G{4}$ are
now independent fields (which we will call black boxes) and are no longer
given by (\ref{berg-G2n}). Note that their Bianchi identities are imposed by
the Lagrange multipliers $\A{9}$, $\A{7}$ and $\A{5}$. The NS-NS
three-form field strength is given by (\ref{berg-G2n}). Note that
the standard action for IIA supergravity can be obtained by integrating
out the dual potentials in (\ref{berg-dualaction}).
\par
The symmetries of the action are similar to those of the democratic
formulation with some small changes. In the supersymmetry transformations
of gravitino and gaugino, the sums now extend only over $n=0,1,2$:
\begin{align}
  \delta_{{\epsilon}} {e}_{{\mu}}{}^{{a}} =
  & \bar{{\epsilon}}{\Gamma}^{{a}} {\psi}_{{\mu}}\, , \notag \\
  \delta_{{\epsilon}} {\psi}_{{\mu}} =
  & \Big( \partial_{{\mu}} +\tfrac{1}{4}
    \not\!{\omega}_{{\mu}}
    +\tfrac{1}{8}\Gamma_{11}\not\!\! {H}_{\mu}
    \Big) {\epsilon}
    +\tfrac{1}{8} e^{{\phi}} \sum_{n=0,1,2} \frac{1}{(2n)!}
    \not \! {G}^{(2n)} {\Gamma}_{{\mu}}
    ({\Gamma}_{11})^{n}{\epsilon}\, , \notag \\
  \delta_{{\epsilon}} B_{\mu\nu} =
  & -2 \, \bar{{\epsilon}} \Gamma_{[\mu}
    \Gamma_{11} {\psi}_{\nu]}\, , \notag \\
     \delta_{{\epsilon}}{\lambda} =
  & \Big( \! \! \not \! \partial \phi
    - \tfrac{1}{12} {\Gamma}_{11} \not\!\! {H} \Big) {\epsilon}
    + \tfrac{1}{4} e^{{\phi}} \sum_{n=0,1,2} \frac{5-2n}{(2n)!}
    \not\! {G}^{(2n)} ({\Gamma}_{11})^{n}
    {\epsilon}\,, \nonumber\\
  \delta_{{\epsilon}}{\phi} =
  & \tfrac{1}{2} \, \bar{{\epsilon}}{\lambda} \,, \nonumber\\
  \delta_\epsilon \textbf{A} =& \textbf{e}^{-B} \wdg \textbf{E}
  \,, \nonumber \\
\delta _{\epsilon} \textbf{G}= & d \textbf{E} + \textbf{G} \wdg \delta
_\epsilon B - H \wdg \textbf{E}
\,,   \notag \\
  \text{with~~}
  & E^{(2n-1)}_{\mu_1 \cdots \mu_{2n-1}} \equiv
    - e^{-\phi} \, \bar{\epsilon} \,
    \Gamma_{[\mu_1\cdots \mu_{2n-2}} \, (\Gamma_{11})^n \,
\Big((2n-1)\psi_{\mu_{2n-1}]} - \tfrac{1}{2} \Gamma_{\mu_{2n-1}]}
\lambda\Big)\,. \label{berg-IIAsusydual}
\end{align}
The transformation of the black boxes $\textbf{G}$ follow from the
requirement that ${\bf e}^{-B} {\bf G}$ transforms in a total derivative.
Here the formal sums
\begin{align}
  {\bf A} = \sum_{n=1}^5 \A{2n-1}\,, \qquad
  {\bf E} = \sum_{n=1}^5 E^{(2n-1)}\,, \qquad
  {\bf G} =\sum_{n=0}^5 G^{(2n)}\,,
\label{berg-AEsum}
\end{align}
have been used. Note that the first formal sum in (\ref{berg-AEsum}) contains
fields, $A^{(1)}$ and $A^{(3)}$, that do not occur in the action. The same
applies to ${\bf G}$, which contains the extra fields $G^{(6)}, G^{(8)}$
and $G^{(10)}$. Although these fields do not occur in the action, one can
nevertheless show that the supersymmetry algebra is realized on them. To
do so one must use the supersymmetry rules of (\ref{berg-IIAsusydual}) and the
equations of motion that follow from the action (\ref{berg-dualaction}).
\par
The gauge symmetries with parameters $\Lambda^{\tNS}$ and $\LAM{2n}$ are
\begin{align}
  \delta_\Lambda B = & \, d \Lambda^{\tNS} \,, \qquad
    \delta_\Lambda {\bf A} =  \, d {\bf L} - \G{0}
\Lambda^{\tNS}
    - d \Lambda^{\tNS} \wdg \textbf{A} \,, \nonumber \\
  \delta_\Lambda {\bf G} = & \, d \Lambda \wdg
  \big({\bf G} - {\bf e}^B \wdg (d {\bf A} +\G{0}) \big)
  + {\bf e}^B \wdg \Lambda \wdg d \G{0} \,.
\label{berg-gaugeA}
\end{align}
Note that, with respect to the R-R gauge symmetry, the ${\bf A}$
potentials transform as a total derivative while the black boxes are
invariant.
\par
Finally, there are $\mathbb{Z}_2$-symmetries, $(-)^{F_L}$ and $I_9\Omega$,
which leave the action invariant. In contrast to the democratic
formulation these two $\mathbb{Z}_2$-symmetries are valid symmetries even
for $G^{(0)}\ne 0$. The $(-)^{F_L}$-symmetry is given by
\begin{align}
  \big\{ \phi, g_{\mu \nu}, B_{\mu \nu} \big\}
  & \rightarrow \big\{ \phi, g_{\mu \nu}, B_{\mu \nu} \big\}\, , \notag \\
  \big\{ \G{2n}_{\mu_1 \cdots \mu_{2n}}, \A{2n-1}_{\mu_1 \cdots \mu_{2n-1}} \big\}
  & \rightarrow - \big\{ \G{2n}_{\mu_1 \cdots \mu_{2n}}, \A{2n-1}_{\mu_1 \cdots \mu_{2n-1}} \big\} \, , \notag \\
  \big\{ \psi_\mu , \lambda, \epsilon \big\}
  & \rightarrow  + \Gamma_{11}
    \big\{ \psi_\mu , -\lambda, \epsilon \big\}\,,
\end{align}
while the second $I_9\Omega$-symmetry reads
\begin{align}
  x^9 & \rightarrow -x^9\, , \notag \\
  \big\{ \phi, g_{\underline{\mu} \nuv},
   B_{\underline{\mu} \nuv} \big\}
  & \rightarrow \big\{ \phi, g_{\underline{\mu} \nuv},
    -B_{\underline{\mu} \nuv} \big\}\, , \notag \\
  \big\{ \G{2n}_{\underline{\mu_1} \cdots \underline{\mu_{2n}}},
    \A{2n-1}_{\underline{\mu_1} \cdots \underline{\mu_{2n-1}}} \big\}
  & \rightarrow (-)^{n+1} \big\{
    \G{2n}_{\underline{\mu_1} \cdots \underline{\mu_{2n}}},
    \A{2n-1}_{\underline{\mu_1} \cdots \underline{\mu_{2n-1}}} \big\}\,,
    \notag \\
  \big\{ \psi_{\underline{\mu}} , \lambda, \epsilon \big\}
  & \rightarrow +  \Gamma^{9} \big\{ \psi_{\underline{\mu}} ,
    -\lambda, \epsilon \big\}\, .
\end{align}

\section*{\begin{center} \Large ADDING THE BRANE ACTIONS \end{center}}

\label{berg-branesusy}  Having established supersymmetry in the bulk, we now
turn to supersymmetry on the brane. As mentioned in the introduction, our
main interest is in one-dimensional orbifold constructions with 8-branes
at the orbifold points. Using the techniques of the three-brane on the
orbifold in five dimensions~\cite{berg-Bergshoeff:2000zn}, we want to
construct an orientifold using a $\mathbb{Z}_2$-symmetry of the bulk
action. On the fixed points we insert brane actions, which will turn out
to be invariant under the reduced ($N=1$) supersymmetry. For the moment
we will not restrict to domain walls (in this case eight-branes) since
our brane analysis is similar for orientifolds of lower dimension. In the
previous section we have seen that our bulk action possesses a number of
symmetries, among which a parity operation. To construct an orientifold,
the relevant $\mathbb{Z}_2$-symmetry must contain parity operations in
the transverse directions. Furthermore, in order to construct a charged
domain wall, we want for a $p$-brane the $(p+1)$-form R-R potential to be
even. For the $8$-brane the $I_9\Omega$ symmetry satisfies the desired
properties. For the other $p$-branes, it would seem natural to use the
$\mathbb{Z}_2$-symmetry
\begin{equation}
  I_{9,8,\ldots ,p+1} \Omega\equiv (I_9\Omega )(I_8\Omega )\cdots (I_{p+1}\Omega
  )\,,
 \label{berg-defI98p}
\end{equation}
where $I_q\Omega $ is the transformation (\ref{berg-I9Omega}) with $9$ replaced
by $q$, and $I_q$ and $\Omega $ commute. However, for some $p$-branes
($p=2,3,6,7$) the corresponding $C^{(p+1)}$ R-R-potential is odd under
this $\mathbb{Z}_2$-symmetry. To obtain the correct parity one must
include an extra $(-)^{F_L}$ transformation in these cases, which also
follows from T-duality~\cite{berg-Dab98}.
This leads for each $p$-brane to the $\mathbb{Z}_2$-symmetry indicated in
Table~\ref{berg-tbl:Z2Op}.

\begin{table}[htb]
\begin{center}
\begin{tabular}{||c|c|c||}
\hline \rule[-1mm]{0mm}{6mm}
$p$       & IIB    & IIA  \\
\hline \rule[-1mm]{0mm}{6mm}
$9$      & $\Omega$   & - \\
\rule[-1mm]{0mm}{6mm}
$8$&-&$I_9\Omega$\\
\rule[-1mm]{0mm}{6mm}
$7$&$(-)^{F_L}I_{9,8}\Omega$&-\\
\rule[-1mm]{0mm}{6mm}
$6$&-&$(-)^{F_L}I_{9,8,7}\Omega $\\
\rule[-1mm]{0mm}{6mm}
$5$&$I_{9,8,\ldots ,6}\Omega$&-\\
\rule[-1mm]{0mm}{6mm}
$4$&-&$I_{9,8,\ldots ,5}\Omega$\\
\rule[-1mm]{0mm}{6mm}
$3$&$(-)^{F_L}I_{9,8,\ldots ,4}\Omega$&-\\
\rule[-1mm]{0mm}{6mm}
$2$&-&$(-)^{F_L}I_{9,8,\ldots ,3}\Omega$\\
\rule[-1mm]{0mm}{6mm}
$1$&$I_{9,8,\ldots ,2}\Omega$&-\\
\rule[-1mm]{0mm}{6mm}
$0$&-&$I_{9,8,\ldots ,1}\Omega$\\
\hline
\end{tabular}
\caption{The $\mathbb{Z}_2$-symmetries used in the orientifold
construction of an O$p$-plane. The T-duality transformation from IIA to
IIB in the lower dimension induces each time a $(-)^{F_L}$. }
\label{berg-tbl:Z2Op}
\end{center}
\end{table}

Thus the correct $\mathbb{Z}_2$-symmetry for a general IIA O$p$-plane is
given by
\begin{equation}
((-)^{F_L})^{p/2} I_{9,8,\ldots ,p+1}\Omega\, . \label{berg-z2}
\end{equation}
The effect of this $\mathbb{Z}_2$-symmetry on the bulk fields reads (the
underlined indices refer to the worldvolume directions, i.e.~$\mu =
(\underline{\mu}, p+1,\ldots ,9)$
\begin{align}
  \big\{ x^{p+1}, \ldots, x^9 \big\}
  & \rightarrow -
  \big\{ x^{p+1}, \ldots, x^9 \big\} \, , \notag \\
  \big\{ \phi, g_{\underline{\mu} \nuv}, B_{\underline{\mu} \nuv} \big\}
  & \rightarrow
  \big\{ \phi, g_{\underline{\mu} \nuv}, -B_{\underline{\mu} \nuv} \big\}\,, \notag \\
  \big\{ \A{5}_{\underline{\mu_1} \cdots \underline{\mu_5}},
    \A{9}_{\underline{\mu_1} \cdots \underline{\mu_9}},
    \G{2}_{\underline{\mu }\nuv}
    \big\}
  & \rightarrow (-)^{\tfrac{p}{2}}
    \big\{
    \A{5}_{\underline{\mu_1} \cdots \underline{\mu_5}},
    \A{9}_{\underline{\mu_1} \cdots \underline{\mu_9}},
    \G{2}_{\underline{\mu }\nuv}
    \big\}\, , \notag \\
  \big\{
    \A{7}_{\underline{\mu_1} \cdots \underline{\mu_7}},
    \G{0}, \G{4}_{\underline{\mu _1}\cdots \underline{\mu _4}}\big\}
  & \rightarrow (-)^{\tfrac{p}{2}+1}
    \big\{
    \A{7}_{\underline{\mu_1} \cdots \underline{\mu_7}},
    \G{0}, \G{4}_{\underline{\mu _1}\cdots \underline{\mu _4}}\big\}\, ,
    \displaybreak[2] \notag \\
  \big\{ \psi_{\underline{\mu}}, \epsilon \big\}
  & \rightarrow
    -\alpha \Gamma^{p+1 \cdots 9} (-\Gamma_{11})^{\tfrac{p}{2}}
    \big\{ \psi_{\underline{\mu}},
    \epsilon \big\}\, , \notag \\
  \big\{ \lambda \big\}
  & \rightarrow + \alpha \Gamma^{p+1 \cdots 9} (+\Gamma_{11})^{\tfrac{p}{2}}
    \big\{ \lambda \big\}\,,
\label{berg-IIAsymmetry}
\end{align}
and for fields with other indices there is an extra minus sign for each
replacement of a worldvolume index $\underline{\mu }$ by an index in a
transverse direction. We have left open the possibility of combining the
symmetry with the sign change of all fermions. This possibility
introduces a number $\alpha =\pm 1$ in the above rules. This symmetry
will be used for the orientifold construction.
\par
For this purpose we choose spacetime to be $\mathcal{M}^{p+1} \times
{T}^{9-p}$ with radii $R^{\overline{\mu}}$ of the torus that may depend
on the world-volume coordinates. All fields satisfy
\begin{equation}
\Phi(x^{\overline{\mu}})=\Phi(x^{\overline{\mu}}+2\pi
R^{\overline{\mu}})\, ,\label{berg-cyclicfields}
\end{equation}
with $\overline{\mu}=(p+1,\ldots,9)$. 
The parity symmetry (\ref{berg-z2}) relates the fields in the bulk  
at $x^{\overline{\mu}}$ and $-x^{\overline{\mu}}$. At the fixed
point of the orientifolds, however, this relation is local and projects
out half the fields. This means that we are left with only $N=1$
supersymmetry on the fixed points, where the branes will be inserted.
Consider for example a nine-dimensional orientifold. The projection
truncates our bulk $N=2$ supersymmetry to $N=1$ on the brane; only half
of the 32 components of $\epsilon$ are even under (\ref{berg-IIAsymmetry}).
The original field content, a $D=10$, $(128 + 128)$, $N=2$ supergravity
multiplet, gets truncated on the brane to a reducible $D=9$, $(64 + 64)$,
$N=1$ theory consisting of a supergravity plus a vector multiplet. One
may further restrict to a constant torus. This particular choice of
spacetime then projects out a $N=1$ $(8+8)$ vector multiplet (containing
$e_{\dot{9}}{}^9$), leaving us with the irreducible $D=9$, $(56 + 56)$,
$N=1$ supergravity multiplet. Similar truncations are possible in lower
dimensional orientifolds, on which the $(64 + 64)$ $N=1$ theory also
consists of a number of multiplets.
\par
We propose the $p$-brane action ($p=0,2,4,6,8$) to be proportional to
\begin{align}
  \mathcal{L}_p = -e^{-\phi} \sqrt{-g_{(p+1)}}
 - \alpha \tfrac{1}{(p+1)!} \varepsilon^{(p+1)} \C{p+1} \,,
\text{~~with~} \varepsilon^{(p+1)} \C{p+1} \equiv
  \varepsilon^{(p+1)}_{\underline{\mu_0} \cdots \underline{\mu_p}} \,
  \C{p+1}{}^{\underline{\mu_0} \cdots \underline{\mu_p}} \,,
\label{berg-braneaction}
\end{align}
with $\varepsilon^{(p+1) \; \underline{\mu_0} \cdots \underline{\mu_p}} =
 \varepsilon^{(10) \; \underline{\mu_0} \cdots \underline{\mu_p}
 \dot{p+1} \cdots \dot{9}}$, which follows from
$e_{\underline{\mu}}{}^{\overline{a}}=0$ (being odd). Here the underlined
indices are $(p+1)$-dimensional and refer to the world-volume. The
parameter $\alpha$ is the same that appears in (\ref{berg-IIAsymmetry}) and
takes the values $\alpha=+1$ for {\it branes}, which are defined to have
tension and charge with the same sign in our conventions, and $\alpha=-1$
for {\it anti-branes}, which are defined to have tension and charge of
opposite signs.  Note that due to the vanishing of $B$ on the brane the
potentials $\C{p+1}$ and $\A{p+1}$ are equal. The $p$-brane action can
easily be shown to be invariant under the appropriate $N=1$ supersymmetry:
\begin{align}
  \delta_\epsilon \mathcal{L}_p =
    - e^{-\phi} \sqrt{-g_{(p+1)}} \,
    \bar{\epsilon} \big( 1- \alpha \Gamma^{p+1 \cdots 9}
    (\Gamma_{11})^{\tfrac{p}{2}} \big) \, \Gamma^{\underline{\mu}} \,
    \big( \psi_{\underline{\mu}} - \tfrac{1}{18} \Gamma_{\underline{\mu}}
    \lambda \big) \, .
\label{berg-delLp}
\end{align}
The above variation vanishes due to the projection under
(\ref{berg-IIAsymmetry}) that selects branes or anti-branes depending on the
sign of $\alpha$ ($+1$ or $-1$ respectively).  In the following
discussions we will assume $\alpha=1$ but the other case just amounts to
replacing branes by anti-branes.
\par
By truncating our theory we are able to construct a brane action that only
consists of bosons and yet is separately supersymmetric. Having these at
our disposal, we can introduce source terms for the various potentials. In
general there are $2^{9-p}$ fixed points. The compactness of the
transverse space implies that the total charge must vanish. Thus the
total action will read (we take the special case that all branes are equally
distributed over all $2^{9-p}$ fixed points)
\begin{align}
  \mathcal{L} = & \mathcal{L}_{\text{bulk}} + k_p \mathcal{L}_p \Delta_p \,,
    \notag \\
  \text{with~~} \Delta_p \equiv &
  \big(\delta(x^{p+1}) - \delta(x^{p+1}- \pi R^{p+1})\big) \cdots
  \big(\delta(x^{9}) - \delta(x^{9}- \pi R^{9})\big)
\label{berg-bulkplbraneaction}
\end{align}
where the branes at all fixed points have a tension and a charge
proportional to $\pm k_p$, a parameter of dimension
$1/\text{[length]}^{p+1}$. Since anti-branes do not satisfy the
supersymmetry condition \eqref{berg-delLp}, we need both positive and negative
tension branes to accomplish vanishing total charge. As explained in the
introduction we are going to interpret the negative tension branes as
O-planes.

The equations of motion following from \eqref{berg-bulkplbraneaction} induce a
$\delta$-function in the Bianchi identity of the $8-p$-form field
strength. In general, an elegant solution is difficult to find, but in 
the eight-brane case the situation simplifies.

\section*{\begin{center} \Large QUANTIZATION OF MASS AND COSMOLOGICAL CONSTANT\end{center}}

Consider the eight-brane case only.
The equation of motion of the nine-form is modified by the brane \& plane
actions such that the solution for $G^{(0)}$ is given by
\begin{equation} \G{0} = \alpha\, \frac{n-8}{2\pi \ell_s}
\varepsilon(x^9)\,. \label{berg-G0}
\end{equation}
Thus we may identify the mass parameter of Type~IIA supergravity as
follows:
\begin{equation}
m= \left\{
  \begin{array}{cc}
\alpha {\displaystyle\frac{n-8}{2\pi\ell_{s}}}\, ,\,\,\,\, & x^9 > 0\, ,\\
& \\
-\alpha {\displaystyle\frac{n-8}{2\pi\ell_{s}}}\, ,\,\,\,\, & x^9 < 0\, .\\
  \end{array}
\right. \label{berg-quant}
\end{equation}
The mass is quantized in string units and it is proportional to $n-8$
where there are $2n$ and $2(16-n)$ D$8$-branes at each O$8$-plane. The
mass vanishes only in the special case $n=8$ when the contribution from
the D$8$-branes cancels exactly the contribution from the O$8$-planes. In
general,  the mass   takes only the
restricted values
\begin{equation}
2\pi\ell_{s} |m| = 0,\, 1,\, 2,\, 3,\, 4,\, 5,\, 6,\, 7,\, 8 .
\label{berg-discrete}
\end{equation}
This is a quantization of our mass parameter, and for the cosmological
constant it follows that
\begin{equation}
m^2= (\G{0})^2 =  \left (\frac{n-8} {2\pi \ell_s} \right )^2\,. \label{berg-CC}
\end{equation}
Thus the mass parameter and the cosmological constant are quantized in
the units of the string length in terms of the integers $n-8$.
\par
The quantization of the mass and of the cosmological constant in $D=10$
was discussed before
in~\cite{berg-Polchinski:1995mt,berg-Polchinski:1996df,berg-Polchinski:1996sm} as well as
in~\cite{berg-Bergshoeff:1996ui,berg-Green:1996bh}.  In the latter two references,
two independent derivations of the quantization condition were given.
In~\cite{berg-Bergshoeff:1996ui}, the T-duality between a $7$-brane \&
$8$-brane solution was investigated. Here it was pointed out that, in the
presence of a cosmological constant, the relation between the $D=10$ IIB
R-R scalar $\C{0}$ and the one reduced to $D=9$, $c^{(0)}$, is given via
a generalized Scherk--Schwarz prescription:
\begin{equation}
\C{0} = c^{(0)} (x^9) + m x^8 \,. \label{berg-axion}
\end{equation}
Here $(x^8, x^9)$ parametrize the 2-dimensional space transverse to the
7-brane. $x^9$ is a radial coordinate whereas $x^8$ is periodically
identified (it corresponds to a U(1) Killing vector field):
\begin{equation}
x^8 \sim x^8 +1 \, . \label{berg-isometry}
\end{equation}
Furthermore, due to the $SL(2,\mathbb{Z})$ U-duality, the R-R scalar
$\C{0}$ is also periodically identified:
\begin{equation}
\C{0} \sim \C{0} +1 \,.
\end{equation}
Combining the two identifications with the reduction rule for $\C{0}$
leads to a quantization condition for $m$ of the form
\begin{equation}
m \sim \frac{n}{\ell_s}\, ,\hskip 2truecm n\,\,\,{\rm integer}\, .
\end{equation}
The same result was obtained by a different method in \cite{berg-Green:1996bh}.

We are able to give a new, and independent, derivation of the
quantization condition for the mass and cosmological constant. The
conditions given in (\ref{berg-quant}), (\ref{berg-CC}) follow straightforwardly
from our construction of the bulk \& brane \& plane action.

Note that the Scherk--Schwarz reduction in (\ref{berg-axion}) and the
quantization of $SL(2,\mathbb{R})$ were essential in deriving the quantization of
$m$. In the new dual formulation we can derive a similar T-duality
relation between the 7-brane and the 8-brane, including the source
terms.  However, in this case the T-duality relation does not imply a
quantization condition for $m$ since we do not know how to realize the
$SL(2,\mathbb{R})$ symmetry in the dual formulation. Another noteworthy feature is
that the derivation of the T-duality rules in the dual formulation does
not require a Scherk-Schwarz reduction. This is possible due to the fact
that the R-R scalar only appears after solving the equations of motion.

\section*{\begin{center} \Large CONCLUSIONS \end{center}}

We have constructed new
formulations of Type~II $D=10$ supergravity.
For both Type~IIA and IIB theories, we constructed democratic bulk
theories with a unified treatment of all R-R potentials.  Due to the
doubling of R-R degrees of freedom one had to impose extra duality
constraints and thus a proper action was not possible. A so-called
pseudo-action, containing kinetic terms for all R-R potentials but
without Chern-Simons terms, was discussed. Furthermore, we have broken
the self-duality explicitly in the IIA case, allowing for a proper
action.  Instead of all R-R potentials only half of the $\C{p}$'s occur
in these theories. Both the standard ($p=1,3$) as well as the dual
($p=5,7,9$) formulations were discussed. Using these actions all bulk \&
brane systems can be described.

A notable difference of our scenario from the HW~\cite{berg-HW,berg-StelleandCo}
scenario is that the walls are the  O$8$ and D$8$ objects which exist in
string theory. The main
goal of the HW theory was to present a scenario for appearance of chiral
fermions starting with $D=11$ supersymmetric theory with non-chiral
fermions. Our O$8$-D$8$ construction may reach this precise goal in an
interesting and controllable way due to stringy nature of this
construction and due to the complete control over supersymmetries in the
bulk \& on the walls. We remark that the strong coupling limit of
Type~I$^\prime$ string theory is equal to the HW theory. Using the
results of this paper, it would be interesting to investigate whether and
how in this limit the O$8$-D$8$ objects can be related to the HW branes.

\section*{\begin{center} \Large ACKNOWLEDGEMENTS \end{center}}

The work described in this talk is based upon the work
of \cite{berg-Bergshoeff:2001pv}. Two of the authors (E.B. and A.V.P.)
would like to thank the organizing committee of the Karpacz School
for the hospitality and stimulating atmosphere provided to us.

This work was supported by the European Commission RTN program
HPRN-CT-2000-00131, in which E.B.  is associated with Utrecht University.
The work of R.K.  was supported by NSF grant PHY-9870115. The work of
T.O.~has been supported in part by the Spanish grant FPA2000-1584.
T.O.~wouldlike to thank the C.E.R.N.~TH Division and the I.T.P.~of the
University of Groningen for their financial support and warm hospitality.
E.B.~would like to thank the I.F.T.-U.A.M./C.S.I.C.~for its hospitality.

\expandafter\ifx\csname natexlab\endcsname\relax\def\natexlab#1{#1}\fi    
                             
\providecommand{\enquote}[1]{``#1''}                                      
                             
\expandafter\ifx\csname url\endcsname\relax                               
                             
  \def\url#1{\texttt{#1}}\fi                                              
                             
\expandafter\ifx\csname urlprefix\endcsname\relax\def\urlprefix{URL }\fi  
                             
\providecommand{\href}[2]{#2}\begingroup\raggedright

\end{document}